\documentclass[sigconf,screen,nonacm] {acmart}
\AtBeginDocument{%
  }

\setcopyright{cc}
\setcctype[4.0]{by}

\begin{document}

\title{Breakdowns for Human-Machine Creative Reflexivity}

\author{Marianne Bossema}
\orcid{1234-5678-9012}
\authornotemark[1]
\affiliation{%
  \institution{University of Applied Sciences}
  \city{Amsterdam}
  \country{The Netherlands}}
\affiliation{%
  \institution{Leiden University}
  \city{Leiden}
  \country{The Netherlands}}

\author{Somaya Ben Allouch}
\affiliation{%
  \institution{University of Applied Sciences}
  \city{Amsterdam}
  \country{The Netherlands}}
\affiliation{%
  \institution{ University of Amsterdam}
  \city{Amsterdam}
  \country{The Netherlands}}

\author{Rob Saunders}
\affiliation{%
  \institution{Leiden University}
  \city{Leiden}
  \country{The Netherlands}}

\renewcommand{\shortauthors}{Bossema et al.}

\begin{abstract}
Generative AI (GenAI) works via goal-directed computation, which differs fundamentally from human creative processes. This poses challenges for the intelligent support of creative experiences. We propose ``breakdowns'' as opportunities for the exchange of perspectives between human and machine. Breakdowns disrupt a flow and force us to consciously evaluate our ``being-in-the-world''. Between human and machine, breakdowns can function as openings for collaborative creative reflection. We are currently studying human-human creative interactions, to identify the markers of these inter-subjective openings, and to understand how they are used in a co-creative process. We present preliminary findings on breakdowns as a design principle for creativity support, prioritising human creative agency and meaningful reflection over automated content generation.
\end{abstract}

\begin{teaserfigure}
  \includegraphics[width=\textwidth]{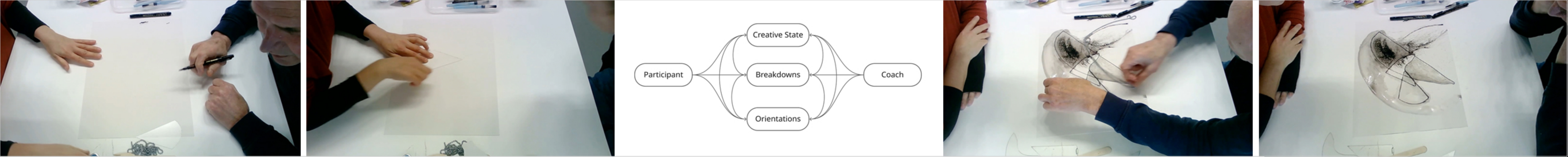}
  \caption{Studying breakdowns in human-human drawing for designing intelligent creativity support}
  \Description{//}
  \label{fig:Teaser}
\end{teaserfigure}

\maketitle

\textbf{Reference:}\newline
Mariannne Bossema, Somaya Ben Allouch and Rob Saunders. 2026. Breakdowns for Human-Machine Creative Reflexivity. In \textit{Proceedings of The First Reflection in Creative Experience (RiCE) Workshop (RiCE W1)}. ACM Creativity \& Cognition 2026, London, UK.

\section{Introduction}
For older adults, creative activities can foster vitality and healthy aging~\cite{cohen2006creativity, fancourt2019evidence}. However, these benefits depend on sustaining personal self-efficacy~\cite{fancourt2019evidence, karwowski2016dynamics}. For this demographic in particular, intelligent creativity support must be able to support creative agency. Rather than steering the process, the system must preserve human autonomy. While GenAI can accelerate creative production, its probabilistic logic is very different from the intuitive, non-linear nature of human creative experience~\cite{bender2021dangers, moruzzi2024user}. Currently, human-ai collaborations are primarily based on explicit descriptions, ignoring the implicit qualities of human creative practice. This results in a lack of context-sensitivity and a detachment from the human experience of making. If creative effort is simply outsourced to GenAI systems, makers become curators of AI-generated content, missing the personal satisfaction and craftsmanship of making~\cite{moruzzi2024user}. However, we can design interactions to support human creative experiences. We previously proposed a framework for intelligent creativity support, based on continuous exchange of human and machine perspectives during creative collaboration~\cite{bossema2025pluri}. We also investigated guiding GenAI to offer real-time alternative explanations, promoting exploration in an embodied creative context~\cite{bossema2026designing}. By continuously following the creative process, an AI agent can achieve greater context-sensitivity, allowing for more meaningful contributions. This raises new questions: How to synchronize with human rhythms of engagement and contribute creatively while preserving human autonomy? 

We propose ``breakdowns'' as a design principle for synchronising human and machine creativity. In ``Being and Time''~\cite{heidegger1962being}, Heidegger describes how a ``breakdown'' occurs when tools or routines fail, disrupting our absorbed, non-reflective engagement with the world. The interruption breaks the ``transparency'' of daily life tools and routines, forcing us to consciously see the object and its context. These moments expose our ``thrownness'', our capability of being immersed in the world and gaining tacit knowledge through experience. Breakdown moments break the immersion and force us to consciously reconsider the situation at hand. In 1987, Winograd and Flores brought these concepts into AI design~\cite{winograd1987understanding}. They argued that technology should be transparent and, in the case of failure, be able to handle the repair of breakdowns. Current intelligent technology enables new forms of collaboration and human-machine participatory sense-making~\cite{de2007participatory}. This calls for re-examining breakdowns in human-machine collaboration. We propose to view breakdowns as meaningful interruptions of creative flow, and as opportunities for collaborative creative reflection. Currently, there is a knowledge gap regarding how breakdowns can function as inter-subjective openings~\cite{guerrero2023forming, tewari2022expecting, wrede2023potentials}. An AI agent that anticipates breakdowns can attune to human rhythms of engagement, and better support creative reflection and exploration. For older adults, such attunement is vital. It has been found that GenAI outputs can overwhelm users~\cite{inuwa2023algorithmic}, and non-experts in particular~\cite{zamfirescu2023johnny}. For a demographic that may feel a loss of control when interacting with autonomous technology~\cite{nimrod2018technostress, zhou2025grand}, breakdowns can be pivotal for preserving and re-assigning creative agency.

\section{Methodology}
We investigate the dynamics of creative breakdowns by first analysing human-human collaborations. We paired 4 creative coaches (experienced visual arts teachers) with 12 older adults for 20-minute 1-to-1 drawing sessions. Each participant was invited to freely explore drawing tools and objects, while a coach facilitated the process. Figure~\ref{fig:Teaser} gives an impression. In this setup, creative coaches acted as facilitators of reflection-in-action~\cite{schon1986reflective}.

\begin{figure*}[hbt!] 
\centering
\includegraphics[width=\textwidth]{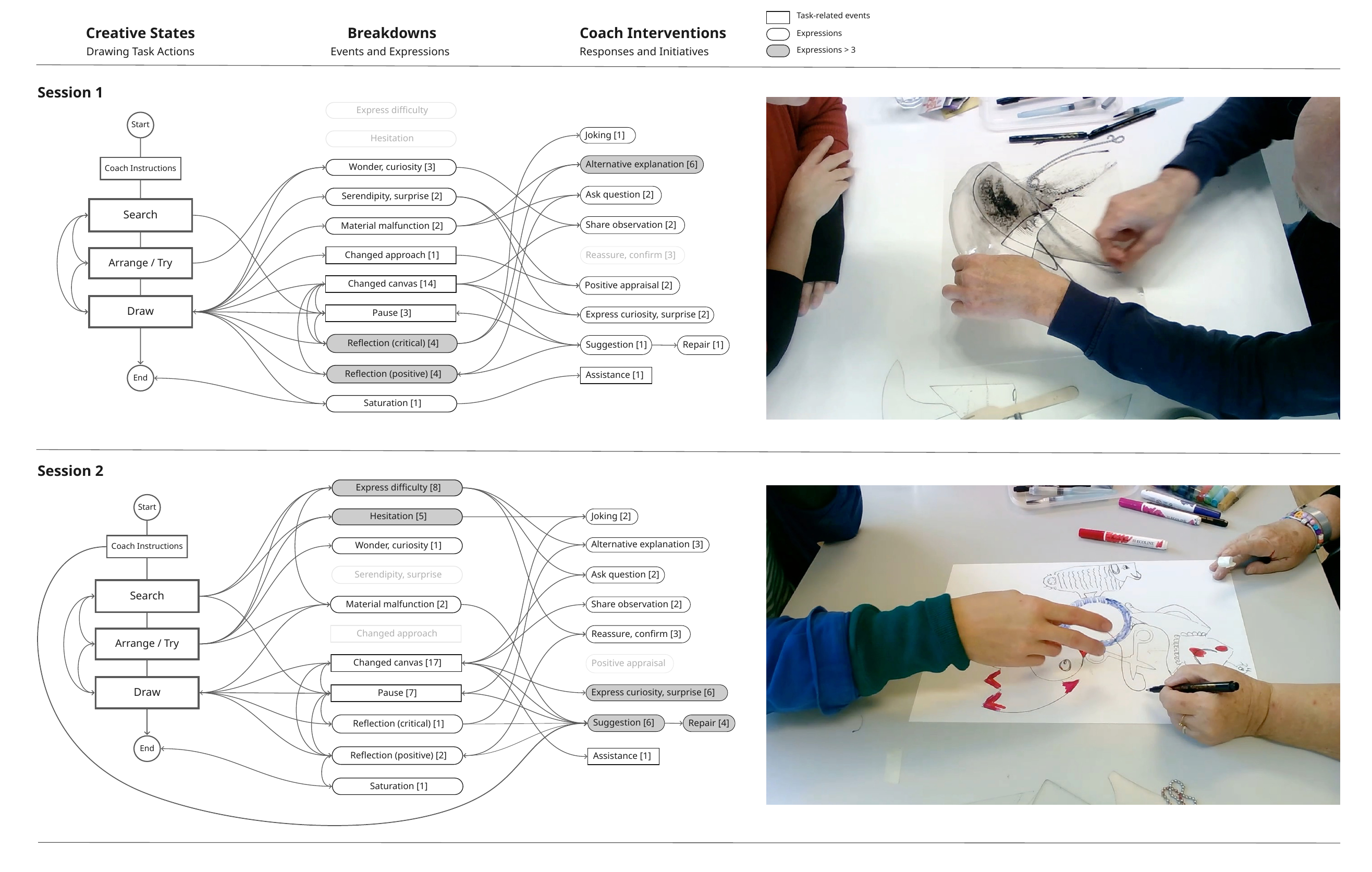} 
\caption{Breakdowns in the context of creative states, eliciting coach interventions, observed in two drawing sessions.}
\Description{//}
\label{fig:Sessions}
\end{figure*}

\begin{figure*} 
\centering
\includegraphics[width=\textwidth]{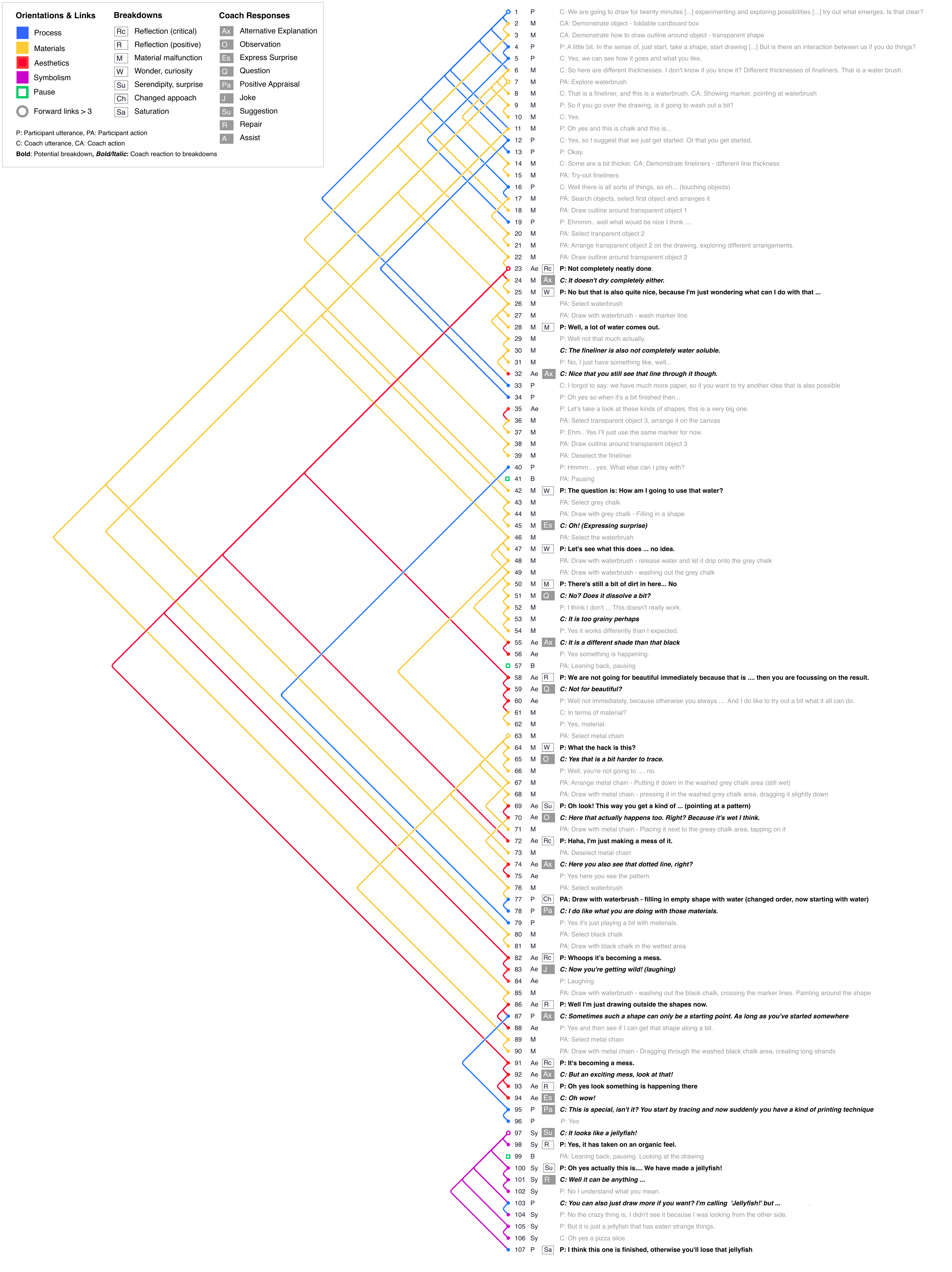} 
\caption{{Linkography visualisation of Session 1, showing orientations, breakdowns and coach interventions.}}
\Description{//}
\label{fig:Linkography}
\end{figure*}

While we have formed a general impression of all 12 sessions, we conducted a detailed analysis of two representative sessions to date. For this analysis, we coded video-recorded data in Atlas.ti~\cite{atlasti2024}, documenting every utterance and action by initiator. We then carried out an in-depth analysis focused on two dimensions: The ``When'': Identifying temporal relationships by mapping creative states (searching, arranging, drawing) against events of breakdowns and coach interventions (Figure~\ref{fig:Sessions}). The ``What'': Tracking conceptual development and alignment by analyzing participant orientations e.g., toward the process, materials, or semiotic aspects of the task (Figure~\ref{fig:Linkography}). 

To analyse the development of orientations within each session, we employed Linkography~\cite{goldschmidt2014linkography}. As defined by Goldschmidt, this method treats creative acts or utterances as a discrete ``moves,'' mapping the relationships between them through backlinks to preceding ideas and forelinks to subsequent events. Linkography is vital to our study because it allows us to visualize how breakdowns align with, or affect orientations. For instance, tracing how moments of impasse link to coach interventions, and how collaborative reflection causes shifts in orientation, opening up new opportunities for exploration. Using Linkoder~\cite{pourmohamadi2011linkographer}, we mapped these temporal and conceptual dependencies (Figure~\ref{fig:Linkography}). We aim to identify a rationale for how an intelligent agent might synchronize and align with, or strategically disrupt human creative processes.

\section{Preliminary Results}

Our initial findings, based on the analysis of two sessions and supported by broader observations of the full dataset, reveal the benefit of dyadic interactions. Although 1-on-1 drawing is less common than solitary practice or group drawing lessons, the dyadic setup promoted creative reflection. The presence of ``the other'' stimulated thinking-aloud, articulating internal dialogues and making them accessible for participatory sense-making~\cite{de2007participatory}. Sessions were very different, but they revealed how coaches navigate between providing creative support and preserving the participant's agency and flow. Coaches were reluctant to take the lead, responding primarily to meaningful events such as breakdowns. This supports our argument that research into breakdowns will benefit the design of AI agents for creativity support.

\subsubsection*{Breakdowns and Creative States}
The data highlights a clear emergence of iterative, non-linear creative states, based on task-related actions: Search, Arrangement \& Try-out, and Drawing (Figure~\ref{fig:Sessions}). We observed that breakdown types are inherently phase-dependent. For instance, Hesitation typically occurred during Search, while Material malfunctions or Serendipity typically occurred while Drawing. Figure~\ref{fig:Sessions} shows that in Session 1, the participant expressed self-critical thoughts (n=4) when concluding a drawing action. This triggered the coach to offer alternative explanations (n=4) as a form of cognitive reappraisal~\cite{brockbank2024cognitive}, prompting the participant to see ``the possible'' in a new light. For example, when the participant expressed, \textit{``It’s becoming a mess,''} the coach responded: \textit{``But an exciting mess, look at that!''}, pointing at emerging patterns in the drawing (Figure~\ref{fig:Linkography}, lines 91, 92). We conclude that breakdowns must be understood within context, by interpreting events given the current creative state.

\subsubsection*{Coach Interventions}
Analysis of coach interventions provides insight into their strategies for creativity support. Although the coaches had very different styles, we observed main types of 1) Affective support, e.g., reassurances and positive appraisal; 2) Attentional support, e.g., sharing observations, expressing curiosity; pointing at specific areas; 3) Creative directions in the form of suggestions and alternative explanations. Because coaches responded to breakdowns, interventions were also phase-dependent, yet they did not always respond. In Session 2 (Figure~\ref{fig:Sessions}), the participant expressed difficulty and hesitation during Search and Arrangement \& Try-out, but the coach responded scarcely. By avoiding interventions during active search, the coach could preserve productive friction and participants' creative agency. During the drawing, the coaches responded to the participant's utterances. Most often, they reacted after completion of drawing actions, when significant changes appeared on the canvas. This contributed to maintaining flow. Furthermore, we identified a coaching strategy of ``reparation''. When coaches offered suggestions, they frequently followed them with neutralising statements, such as: \textit{``Well it can be anything...''} (Figure~\ref{fig:Linkography}, line 101). Coaches used ``strategic silence'', and were cautious in steering the process. Breakdowns functioned as openings for timely, context-sensitive intervention.

\subsubsection*{Breakdowns and Orientations}
Our Linkography analysis suggests that understanding orientations is a crucial for context-sensitive creativity support. Session 1 showed shifts in orientation during the creative task (Figure~\ref{fig:Linkography}). Coach instructions provided an initial orientation to the Process (the approach), after which the focus shifted to Materials (behaviors and affordances), followed by Aesthetics and Symbolism. While trajectories differ between sessions, the data suggests that breakdowns, coach interventions and orientation shifts are interrelated. Moments of impasse can introduce new orientations. Coaches interventions can cause breakdowns through alternative explanations or suggestions, leading to shifts. For instance, toward the end of Session 1, the coach introduced a shift towards Symbolism by sharing the figurative association of a \textit{``jellyfish.''} (Figure~\ref{fig:Linkography}, line 97). The participant then paused, adopted the suggestion, and chose to stop with this drawing \textit{``not to lose that jellyfish''} (Figure~\ref{fig:Linkography}, line 107). This shows the causal relationship of an intervention, orientation shift and breakdown. Coach interventions mostly aligned with the current orientation, allowing participants to maintain the balance between broadening and deepening their creative exploration.

\section{Implications \& Future Directions}
Our findings demonstrate that breakdowns can serve as openings for exchange and reflection, provided they are understood within the context of creative states and orientations. We observed how human coaches used ``strategic silence'' to preserve creative flow and agency, and responded to participants' expressions of breakdowns. While coaches offered creative direction through alternative explanations and suggestions, they remained reluctant to take the lead, using reparations to return agency to the participant.

\subsubsection*{Future Work}
We will further expand our analysis by processing the remaining sessions, to gather a wider range of examples. For instance, some sessions involved joint drawing between coach and participant, and we need to further study the various ways in which coaches supported and contributed creatively. Future research will also investigate automated detection of breakdowns. Few-shot prompting~\cite{liu2023pre} can be used to provide a Vision-Language Model with representative examples from the human-human sessions. We will test whether this improves the model's capacity to reason about creative interventions. In addition, we need to address technical challenges of dynamic state management~\cite{wong2026state}. We plan to use structured grammar and knowledge graphs, which has been shown to support reasoning in continuous interactions~\cite{liang2024survey}. Finally, we will evaluate a functional prototype in human-machine sessions to assess the interaction design using breakdowns for creative reflection to support human creative experiences.

\subsubsection*{Concluding Remarks}
Our findings suggest that attunement can be dynamically negotiated through breakdowns, acting as openings for creative reflection. Human coaches were cautious to interfere, responding primarily to breakdown moments, preserving participants' creative agency. Intelligent creativity support tools can use breakdowns as openings for intervention, but must interpret these moments in context, given the current creative state and orientation. Redefining breakdowns as opportunities for co-creative reflection advances research into intelligent creativity support. However, further research is needed to model breakdowns in the context of human creative experiences, embed these models into interactive prototypes, and evaluate their effectiveness.

\begin{acks}
This work is part of the project Social Robotics and Generative AI to support Creative Experiences for Older Adults (NWO Doctoral Grant for Teachers, project no. 023.019.021).
\end{acks}

\newpage
\bibliographystyle{ACM-Reference-Format}
\bibliography{main}

\end{document}